\documentclass[aps,pra,preprint,groupedaddress,showpacs]{revtex4}
\usepackage[dvips]{graphicx}

\begin{document}
\title{S-wave quantum entanglement in a harmonic trap}
\author{Jia Wang, C. K. Law and M.-C. Chu}
\affiliation{Department of Physics, The Chinese University of Hong
Kong, Shatin, Hong Kong SAR, China}
\date{\today}
\begin{abstract}
We analyze the quantum entanglement between two interacting atoms
trapped in a spherical harmonic potential.  At ultra-cold
temperature, ground state entanglement is generated by the
dominated s-wave interaction. Based on a regularized
pseudo-potential Hamiltonian, we examine the quantum entanglement
by performing the Schmidt decomposition of low-energy
eigenfunctions. We indicate how the atoms are paired and quantify
the entanglement as a function of a modified s-wave scattering
length inside the trap.

\end{abstract}
 \maketitle

\section{Introduction}

The interactions between trapped ultra-cold atoms govern many
interesting collective quantum phenomena, ranging from
Bose-Einstein condensation \cite{Leggett} to recently observed
fermion superfluid \cite{bcs}. At sufficiently low energies, it is
known that short ranged atom-atom interactions can be replaced by
a pointlike regularized pseudo-potential under the shape
independent approximation \cite{Huang}. The strength of such a
pseudo-potential is determined by a s-wave scattering length $a$
only, and hence one can control atom-atom interaction by tuning
the scattering length via the technique of Feshbach resonance. For
trapped systems, the theory of pseudo-potential has been examined
in details by several authors \cite{Busch,Block}. As long as the
range of the actual atom-atom interaction is short compared with
the width of the trap, low energy eigenfunctions can be accurately
captured by the eigenfunctions of the pseudo-potential, except for
a few tightly bound states that may exist inside the range of the
inter-atomic potential. Such a universal applicability is the
essence of shape independent approximation. Therefore the study of
eigenfunctions of pseudo-potentials would provide useful insight
about generic features of two-body correlations in the low energy
regime.

In this paper, we address a fundamental question of how the
scattering length controls quantum correlations between two
ultra-cold atoms inside a harmonic trap. Quantum control of
trapped ultracold atoms has been a subject of considerable
research interest, regarding potential applications in quantum
information processing \cite{jaksch1,jaksch2}. For example,
collisions of atoms can be exploited to perform various quantum
logic operations \cite{jaksch2}. However, the nature of quantum
entanglement arising from s-wave scattering has not been fully
explored \cite{frees}. Such an entanglement is inherent in the
continuous degree of freedom of atoms, and it may have effects on
the fidelity of quantum gates based upon collisional mechanisms
\cite{jaksch2}. In this paper we will analyze the quantum
entanglement of the low energy eigenstates defined by the
regularized pseudo-potential and the harmonic trap. By performing
the Schmidt decomposition of low energy eigenfunctions, we show
that quantum entanglement is manifested as pairing of atoms in a
set of orthogonal mode functions in three-dimensional space. In
particular, the angular momenta are identified as good quantum
numbers to characterize the Schmidt mode functions. We will
present numerical results that quantify the degree of entanglement
as a function of the scattering length $a$. In addition, we will
examine the entanglement in the $a \to \infty$ limit. Such a
strong coupling regime corresponds to the unitarity limit in
degenerate quantum gases \cite{ho}. The study of pairing in
two-body models in such a limit may shed light on quantum
correlations in the more difficult many body problems.

\section{Regularized Hamiltonian and energy eigenstates}

To begin with, we consider a system of two interacting atoms with
equal mass trapped in a spherical harmonic potential. The
Hamiltonian of the system is given by,
\begin{equation}
H =  - \frac{{\hbar ^2 }}{{2m}}\nabla _1^2  - \frac{{\hbar ^2
}}{{2m}}\nabla _2^2  + \frac{1}{2}m\omega ^2 r_1^2  +
\frac{1}{2}m\omega ^2 r_2^2  + V ( {\bf r}_1 -{\bf r}_2),
\end{equation}
where ${\bf r}_1$ and ${\bf r}_2$ are the position vectors of the
two atoms, $m$ is the mass of the atom, and $\omega$ is the trap
frequency. The interaction between the two atoms is described by
the short range potential $V$ which will be replaced by a
pseudo-potential in Eq. (3) under the shape independent
approximation. For convenience, we separate the Hamiltonian into a
center-of-mass part and a relative part, $H = H_{cm}  + H_{rel}$
so that
\begin{eqnarray}\label{eq1}
&& H_{cm}  =  - \frac{1}{8}\nabla _R^2  + 2R^2, \\ && H_{rel} =
-\frac{1}{2}\nabla _r^2  + \frac{1}{2}r^2 + 2\pi a \delta ^{\left(
3 \right)} \left( \mathbf{r} \right)\frac{\partial }{{\partial
r}}r,
\end{eqnarray}
with ${\bf  R} = ({\bf r}_1  + {\bf r}_2 )/2$ and ${\bf r} = {\bf
r}_1  - {\bf r}_2$. Here the energy and length are expressed in
units of $\hbar \omega$ and $\left( {\hbar /\mu \omega }
\right)^{1/2}$ (where $\mu=m/2$ is the reduced mass) respectively.
The strength of the pseudo-potential is characterized by the
modified s-wave scattering length $a$. For a given inter-atomic
potential, the precise value of $a$ depends on the trapping
potential and it can be determined self-consistently by the
methods discussed in Ref. \cite{Block}. In this paper we will
treat $a$ as a parameter of the Hamiltonian.

The s-wave eigen-functions of the $H$ have been solved
analytically in Ref. \cite{Busch}. Given a scattering length, the
eigen-energy $E$ of $H_{rel}$ is defined by the solution of:
\begin{equation}\label{eq3}
\frac{{\Gamma \left( { - E/2  - 1/4} \right)}}{{2\Gamma \left( { -
E/2 +3/4 } \right)}} = a
\end{equation}
with $\Gamma$ being the gamma function. The eigenfunctions of
$H_{rel}$ with the energy $E$ are given by:
\begin{eqnarray}
\psi_E (\mathbf{r}) = A e^{ - r^2 /2}  U \left(-E/2+3/4 , 3/2, r^2
\right)
\end{eqnarray}
where $A$ is a normalization constant, and $U(\alpha,\beta,z)$ is
the Kummer's function  \cite{Handbook}.

In this paper we assume that the center of mass wave function is
the ground state of $H_{cm}$, which is a simple gaussian: $\Phi (
\mathbf{R} ) = {{2\sqrt 2 }}e^{ - 2R^2 }/{{\pi ^{3/4} }}$.
Combining $\Phi ( \mathbf{R} )$ with $\psi_E (\mathbf{r})$, the
two-particle energy eigenfunctions are given by
\begin{equation}\label{eq6}
\Psi \left( {{\bf r}_1,{\bf r}_2 } \right) = \Phi \left(
\mathbf{R} \right)\psi_E \left( \mathbf{r} \right).
\end{equation}
In Fig. 1, we illustrate how the eigen-energies depend on the
scattering length. For convenience, we choose to plot with the
inverse of scattering length, i.e., $1/a$, in order to indicate
the continuous branch associated with the ground states. It should
be noted that the ground state energy becomes large and negative
when $1/a$ is positive and large. This feature also occurs in the
absence of the harmonic trapping potential, and it is due to the
existence of a tightly bound state in the $a \to 0^+$ limit
\cite{Busch}. As an illustration, we show in Fig. 2 the radial
probability density associated with relative coordinate wave
functions at several values of $a$.

\section{Schmidt Decomposition }

The characterization of quantum entanglement is achieved by
Schmidt decomposition of $\Psi \left( {{\bf r}_1 ,{\bf r}_2 }
\right)$, which reads:
\begin{equation}
\Psi \left( {{\bf r}_1,{\bf r}_2} \right) = \sum\limits_{j} {\sqrt
{\lambda _j } u_j \left( {\mathbf{r}_1 } \right)v_j \left(
{\mathbf{r}_2 } \right)},
\end{equation}
where $\lambda _j$ are eigenvalues, and $u _j$ and $v _j$ are
Schmidt eigenmodes defined by,
\begin{eqnarray}
&& \int\limits_{} {} d{\bf r}_1'\int\limits_{} {d {\bf r}_2 } \Psi
\left( {{\bf r}_1,{\bf r}_2} \right) \Psi^* \left( {{\bf
r}_1',{\bf r}_2} \right) u_j ({\bf r}_1') = \lambda _j u_j({\bf
r}_1),
\\ &&
\int\limits_{} {} d{\bf r}_2'\int\limits_{} {d {\bf r}_1 } \Psi
\left( {{\bf r}_1,{\bf r}_2} \right) \Psi^* \left( {{\bf r}_1,{\bf
r}_2'} \right) v_j ({\bf r}_2') = \lambda _j v_j ({\bf r}_2).
\end{eqnarray}
Note that the mode functions $u_j$ form a complete and orthonormal
set, and the same is true for $v_j$. If the atom 1 appears in the
mode $u_j$, then with certainty the atom 2 must be in the mode
$v_j$. In other words, Eq. (7) indicates the pairing structure of
the two particle state. In addition, the distribution of
$\lambda_j$ provides a measure of the degree of entanglement. This
is usually discussed in terms of the entanglement entropy $S = -
\sum\nolimits_j {\lambda _j } \log \lambda _j$. However, a more
transparent measure is the effective number of Schmidt modes,
which is provided by the Schmidt number: $K \equiv
1/\sum\nolimits_j {\lambda _j^2 }$ \cite{grobe}. A disentangled
(product) state corresponds to $K=1$, i.e., there is only one term
in the Schmidt decomposition. The larger the value of $K$, the
higher the entanglement. We point out that $1/K$ equals the purity
of the density matrix of an individual particle. The purity has
also been employed as a measure of the degree of entanglement in
various physical situations \cite{purity}.

To carry out the Schmidt decomposition of the wave functions in
three dimension, we note that $R = \sqrt {r_1^2 + r_2^2 + 2r_1 r_2
\cos \gamma } /2$ and $r = \sqrt {r_1^2 + r_2^2 - 2r_1 r_2 \cos
\gamma }$, where $\gamma$ is the angle between $\mathbf{r}_1$ and
$\mathbf{r}_2$. Therefore the wave function $\Psi \left( {{\bf
r}_1,{\bf r}_2} \right) = \Psi \left( {r_1 ,r_2 ,\cos \gamma }
\right) = \sum\limits_{l = 0}^\infty {\alpha _l \left( {r_1 ,r_2 }
\right)P_l \left( {\cos \gamma } \right)}$, where $P_l (x)$ is the
Legendre polynominal, and
\begin{equation}
\alpha _l \left( {r_1 ,r_2 } \right) = \frac{{2l +
1}}{2}\int_0^\pi  {d\gamma \sin \gamma  \cdot \Psi \left( {r_1
,r_2 ,\cos \gamma } \right)P_l \left( {\cos \gamma } \right)}.
\end{equation}
With the help of the addition formula: $(2l + 1)P_l \left( {\cos
\gamma } \right) = 4\pi \sum\limits_{m =  - l}^l {Y_{lm}^{*}
\left( {\theta _1 ,\phi _1 } \right)Y_{lm} \left( {\theta _2 ,\phi
_2 } \right)}$, we have
\begin{equation}
 \Psi ( {\bf r}_1, {\bf r}_2) = 4 \pi \sum\limits_{l = 0}^\infty
\sum\limits_{m =  - l}^l { \frac{\alpha _l \left( {r_1 ,r_2 }
\right)} {2l+1} Y^{*}_{lm} \left( {\theta _1 ,\phi _1 }
\right)Y_{lm} \left( {\theta _2 ,\phi _2 } \right) }.
\end{equation}
This expression is already partially in the Schmidt form, because
the pairing of angular functions has been identified. The
remaining task is to decompose $\alpha _l \left( {r_1 ,r_2 }
\right)$ for each $l$. This can be achieved by performing the
Schmidt decomposition of the function $r_1r_2\alpha_l(r_1,r_2)$,
i.e.,
\begin{equation}
r_1 r_2 \alpha _l \left( {r_1 ,r_2 } \right) = \sum\limits_n
{\sqrt {\lambda _{nl} }  u_{nl} \left( {r_1 } \right) v_{nl}
\left( {r_2 } \right)}.
\end{equation}
Here the prefactor $r_1r_2$ is introduced in order to ensure
correct normalization in radial directions.

The final form of the Schmidt decomposition of the wave function
Eq. (5) now reads:
\begin{equation}
 \Psi \left( {{\bf r}_1,{\bf r}_2} \right)= \sum\limits_{n =
1}^\infty  {\sum\limits_{l = 0}^\infty  {\sum\limits_{m =  - l}^{
l} {\left( { \frac{{4\pi \sqrt{\lambda _{nl}}}}{{2l + 1}}}
\right)} } }\left[ { \frac {{u_{nl} (r_1)}} {{r_1}} Y_{lm}^{*}
\left( {\theta _1 ,\phi _1 } \right)} \right]\left[ { \frac
{{v_{nl} (r_2)}} {{r_2}} Y_{lm} \left( {\theta _2 ,\phi _2 }
\right)} \right].
\end{equation}
Our derivation indicates a general feature that for any wave
function that is a function of distances $R$ and $r$ only, the
angular parts of Schmidt modes are simply the spherical harmonics.
The pairing involves angular momentum quantum numbers $(l,m)$ and
$(l,-m)$, i.e., the same $l$ and opposite $m$ must be paired. In
addition, each $m$ is of equal weight for a given $l$. Therefore
if one could select Schmidt modes with a fixed $n$ and $l$ via
projective measurement, then the resulting state is a maximally
entangled state among various $m$'s on the projected $l$ manifold.

\section{Numerical Results}

The fact that the angular parts of Schmidt modes can be obtained
analytically reduces the computational difficulty in the original
three-dimensional problem. Finding $u_{nl}$, $v_{nl}$ and $\lambda
_{nl}$ in Eq. (12) is a relatively simple numerical task, because
$r_1r_2\alpha_l(r_1,r_2)$ behaves as a two-particle (one
dimensional) wave function in non-negative $r_1$ and $r_2$
regions. Specifically, we first obtain $\alpha_l(r_1,r_2)$ by
performing numerical integration of (10) for discretized values of
$r_1$ and $r_2$, typically with the spacing $\Delta r=0.01$. For
the s-wave eigenfunctions considered here, it is sufficient to
choose $r_1$ and $r_2$ ranging from $0$ to $3.5$, where the wave
functions are mainly confined.  Because of the symmetry properties
of $r_1r_2\alpha_l(r_1,r_2)$, $u_{nl}$ and $v_{nl}$ are the same
real functions. Therefore  $u_{nl}$, $v_{nl}$ and $\lambda _{nl}$
can be obtained by diagonalizing the matrix
$r_1r_2\alpha_l(r_1,r_2)$. For low energy states considered in
this paper, $l$ up to 30 are typically sufficient in order to
obtain convergent results.

We can now obtain the value of $K$ for the state (13) from the
Schmidt eigenvalues, i.e.,
\begin{equation}
K =  1/ \sum\limits_{n = 1}^\infty  {\sum\limits_{l = 0}^\infty {
{\Lambda _{nl}^2  } } }.
\end{equation}
Here $\Lambda _{nl} = 16\pi ^2 \lambda _{nl} /(2l + 1)^{3/2}$ is
defined. The main result of this paper is shown in Fig. 3, where
the values of $K$ are displayed as a function of the inverse of
the scaled scattering length for low energy states. We notice that
higher excited states generally have higher quantum entanglement.
However, the ground state (curve A) shows a distinct behavior in
the positive $1/a$ region, where we notice a sharp rise of $K$ as
$1/a$ increases.

The strong ground state entanglement in the large positive $1/a$
limit is understood as the appearance of increasingly bounded
atoms. This is implied in the ground state energy curve (A) in
Fig. 1, as well as in the wave function $(1/a=2)$ shown in Fig. 2.
A crude estimation of $K$ can be made from the analytical results
of gaussian functions. For gaussian functions separable in the
center of mass and relative coordinates, it is known that $K
\propto (\Delta R)^3 /(\Delta r)^3$ when the center of mass width
$\Delta R$ is much wider than the width of the relative coordinate
$\Delta r$ \cite{fedorov}. Here the tightly bound state
corresponds to a strong localization in particle's relative
distance, with $\Delta r$ of order $a$ in the $1/a \gg 1$ limit.
Therefore $\Delta R /\Delta r \gg 1 $, and hence high values of
$K$ can be expected as in the case of gaussians functions.
However, we remark that gaussian can only serve as a guide here,
because the singular $1/r$ dependence in $\psi_E (r)$ cannot be
captured by gaussians. Indeed, $K$ depends on $a$ in a complicated
form according to our numerical calculations.

For excited eigenstates on the branches B and C, we see an
interesting feature in Fig. 3 that the change of quantum
entanglement is only sensitive to a range of scaled scattering
lengths. Such a window of scaled scattering lengths is highlighted
by dash lines in Fig. 3, where we found that $K$ changes
significantly with $1/a$ when $1/|a|<2$. Recalling that we are
using the length unit defined by the harmonic trap, our results
suggest that the s-wave interaction can appreciably affect quantum
entanglement in excited states when $a$ is greater than or
comparable with the width of a ground state particle inside the
trap, i.e., $|a| > \left( {\hbar /2m \omega } \right)^{1/2}$. This
feature is also true for the ground state with negative scattering
lengths.

The structure of quantum entanglement is characterized by the
distribution of Schmidt eigenvalues and the Schmidt mode
functions. In Fig. 4 we show the distribution $\Lambda_{nl}$ for
the ground state with $1/a=-2$ and $1/a=2$. In the case of
$1/a=-2$, the entanglement $K=1.2$ is not high because of the
presence of a dominant $(n,l,m)=(1,0,0)$ Schmidt mode, which
covers about $70\%$ probability of the state. On the contrary,
$K=10.13$ is much higher for the $1/a=2$ case. This is shown in
the distribution of $\Lambda_{nl}$ (Fig. 5b) in which Schmidt
modes with $n=1$ and higher angular momentum number $l$ are
involved. In other words, the strong entanglement is mainly
manifested in the angular variables.

In Fig. 5, we illustrate the shape of leading Schmidt modes
corresponding to the ground state wave functions in Fig. 4. Since
the angular part are simple spherical harmonic functions, we show
only the radial mode functions in the figure. Apart from the fact
that Schmidt modes of positive scattering lengths are more
attracted to the origin, the change of scattering length has small
effects on the mode functions. This is in contrast to the Schmidt
eigenvalues, which are more sensitive to the values of $a$ as
depicted in Fig. 4.

Finally, let us discuss the limit of $|a| \to \infty$ or $1/|a|
\to 0$. In such a limit, we find that the eigenfunctions take
simple analytic forms. We present some of the s-wave
eigenfunctions of $H$ in the limit $|a| \to \infty$:
\begin{eqnarray}
&& \Psi_{10} \left( {{\bf r}_1 ,{\bf r}_2 } \right) = \frac{{ 2
}}{{\pi ^{3/2} }}\frac{{e^{ - r_1^2 } e^{ - r_2^2 } }}{{\left|
{{\bf r}_1 - {\bf r}_2 } \right|}},  \\ && \Psi_{11} \left( {{\bf
r}_1 ,{\bf r}_2 } \right) = \frac{{\sqrt 2 }}{{\pi ^{3/2}
}}\frac{{e^{ - r_1^2 } e^{ - r_2^2 } }}{{\left| {{\bf r}_1 - {\bf
r}_2 } \right|}}\left[ {1 - \left( {{\bf r}_1 - {\bf r}_2 }
\right)^2 } \right],
\\
&& \Psi_{12} \left( {{\bf r}_1 ,{\bf r}_2 } \right) = \frac{{\sqrt
{3/2} }}{{\pi ^{3/2} }}\frac{{e^{ - r_1^2 } e^{ - r_2^2 }
}}{{\left| {{\bf r}_1 - {\bf r}_2 } \right|}}\left[ {1 - 4\left(
{{\bf r}_1 - {\bf r}_2 } \right)^2 + \frac{4}{3}\left( {{\bf r}_1
- {\bf r}_2 } \right)^4 } \right].
\end{eqnarray}
Here Eqs. (15-17) correspond to the energy eigenfunctions (center
of mass $+$ relative coordinates) associated with the lowest three
states of $H_{rel}$, and the ground state of $H_{cm}$. Although
Eqs. (15-17) are simple expressions, the Schmidt decomposition
still cannot be carried out analytically. We perform numerical
calculations which give: $K_{10} = 1.98$, $K_{11} =3.45$, and
$K_{12}=9.11$ for the wave functions given in Eqs. (15-17).
Therefore the degree of entanglement remains finite at the
infinite scattering limit.

We remark that the regularized pseudo-potential method can only
describe wave functions at inter-atomic distance much larger than
the range of the actual inter-atomic potential $b$, i.e., $|{\bf
r}_1-{\bf r}_2| \gg b$. Inside the interaction range $b$, the
actual wave functions remain finite as $|{\bf r}_1-{\bf r}_2| \to
0$. Therefore the singular behavior at ${\bf r}_1={\bf r}_2 $ in
Eqs. (15-17) is only an artifact of the shape independent
approximation, and hence these wave functions should be understood
for $|{\bf r}_1-{\bf r}_2| \gg b$ only. However, since $b$ is
typically much smaller than $\left( {\hbar /\mu \omega }
\right)^{1/2}$ (which is about the width of $\Psi$), the
probability of finding the two particles within the range $b$ is
negligible \cite{number}. This justifies the use of the shape
independent approximation here.

\section{Summary}

To summarize, we present a procedure to analyze the s-wave quantum
entanglement between two ultra-cold atoms in a spherical harmonic
trap. The s-wave interaction is described by the regularized
pseudo-potential. By performing the Schmidt decomposition of low
energy eigenstates, we quantify the quantum entanglement and
indicate its dependence on the modified scattering length. In
particular, our Schmidt analysis reveals the angular correlations
by showing explicitly the pairing of spherical harmonic functions.
For small and positive $a$, the ground states are highly entangled
states, and we explain this feature as a consequence of tight
binding between the atoms. For low excited states and ground
states with negative $a$, we find that the atom-atom interaction
can only appreciably affect the entanglement when the scattering
length is larger than the width of the (non-interacting) ground
state defined by the trap. However, the degree of entanglement
remains finite in the large scattering length limit.  Our work
here indicates that quantum entanglement can be controlled by the
scattering length. To explore applications of s-wave entanglement,
one may need to establish schemes for detecting Schmidt modes. In
addition, the dynamics of entanglement associated with
non-stationary states of the system is also an interesting topic
for open future investigations.

\begin{acknowledgments}
The authors thank Stephen K. Y. Ho for discussions. This work is
supported in part by the Research Grants Council of the Hong Kong
Special Administrative Region, China (Project No. 400504).
\end{acknowledgments}

\newpage
\begin{figure}
\includegraphics[width=10.0cm]{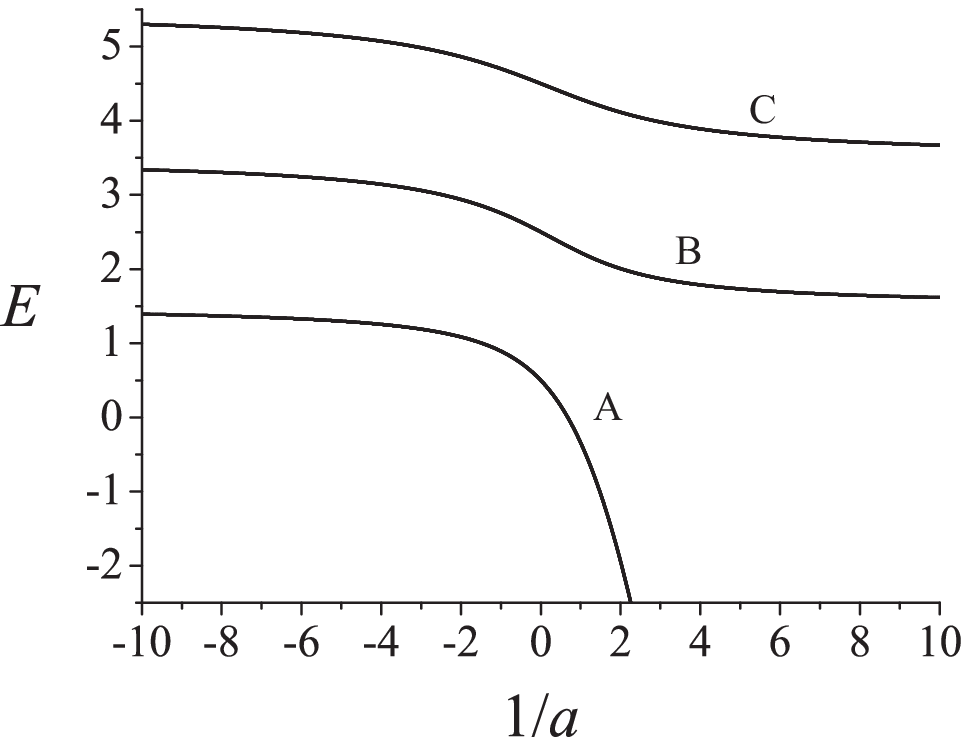}
\caption{Eigen-energies of $H_{rel}$ as a function of the inverse
of dimensionless scaled scattering length. $E$ is in units of
$\hbar \omega$. The three branches: A, B, and C correspond to the
lowest three states with zero angular momentum (in relative
coordinate).}
\end{figure}

\begin{figure}
\includegraphics[width=10.0cm]{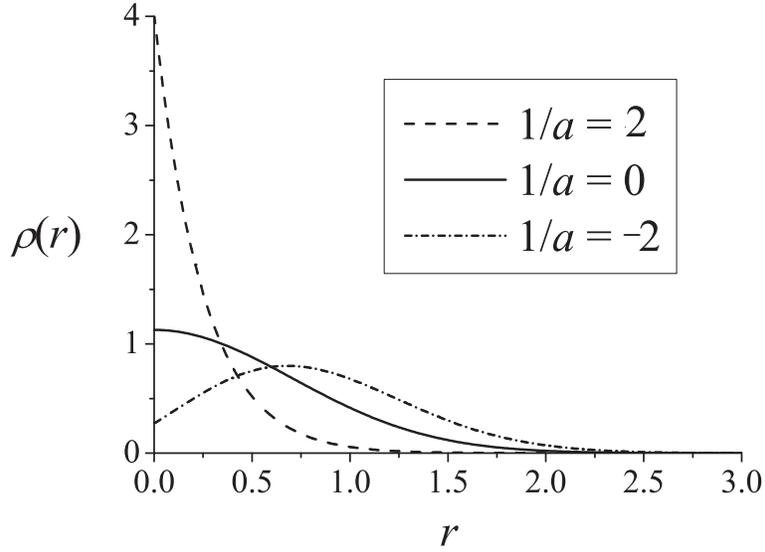}
\caption {Radial densities $\rho (r) \equiv 4 \pi r^2
|\psi_E(r)|^2 $ associated with the ground state wave functions at
$1/a=0, \pm 2$.  The radial distance $r$ is in units of $\left(
{\hbar /\mu \omega } \right)^{1/2}$.}
\end{figure}

\begin{figure}
\includegraphics[width=10.0cm]{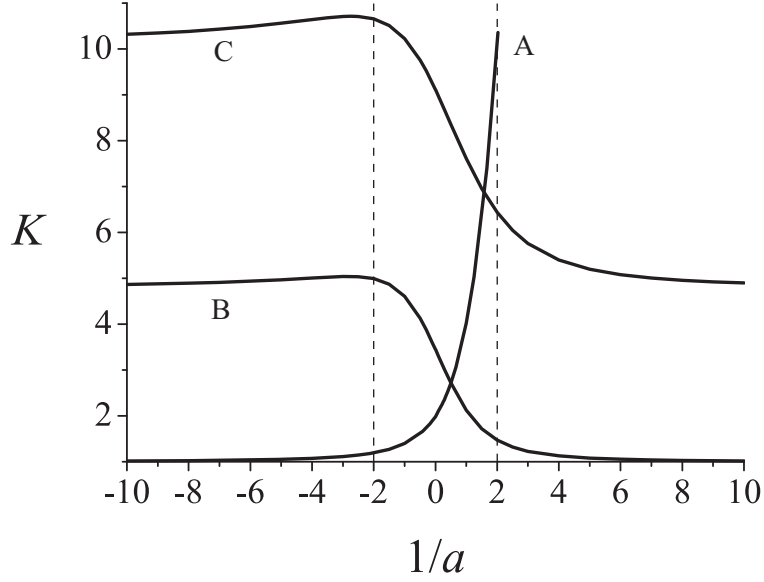}
\caption{Schmidt number as a function of the inverse of
dimensionless scaled scattering length for the low eigenstates
associated with the three energy curves in Fig. 1. $K$ is plotted
up to $1/a=2$ for the ground state curve A because the high values
of $K$ is out of the range of the figure for $1/a >2$.}
\end{figure}

\begin{figure}
\includegraphics[width=12.0cm]{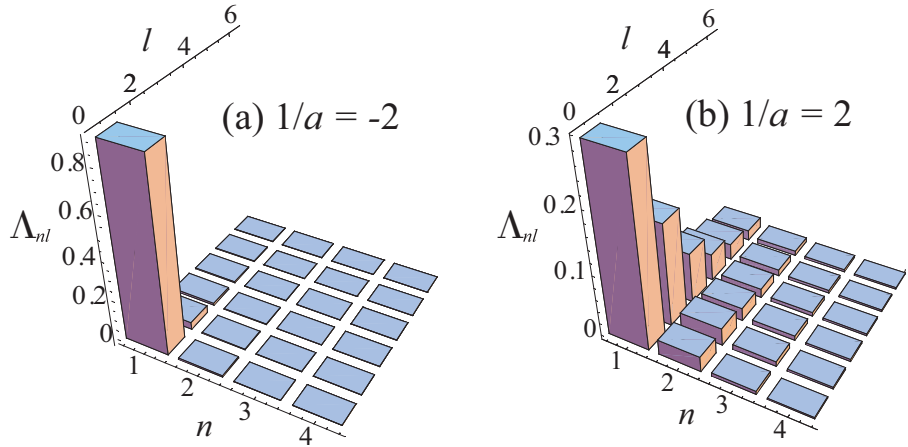}
\caption{(Color online) Distribution of $\Lambda_{nl}$ (see text)
of ground states for: (a) $1/a=-2$ and (b) $1/a=2$.}
\end{figure}

\begin{figure}
\includegraphics[width=12.0cm]{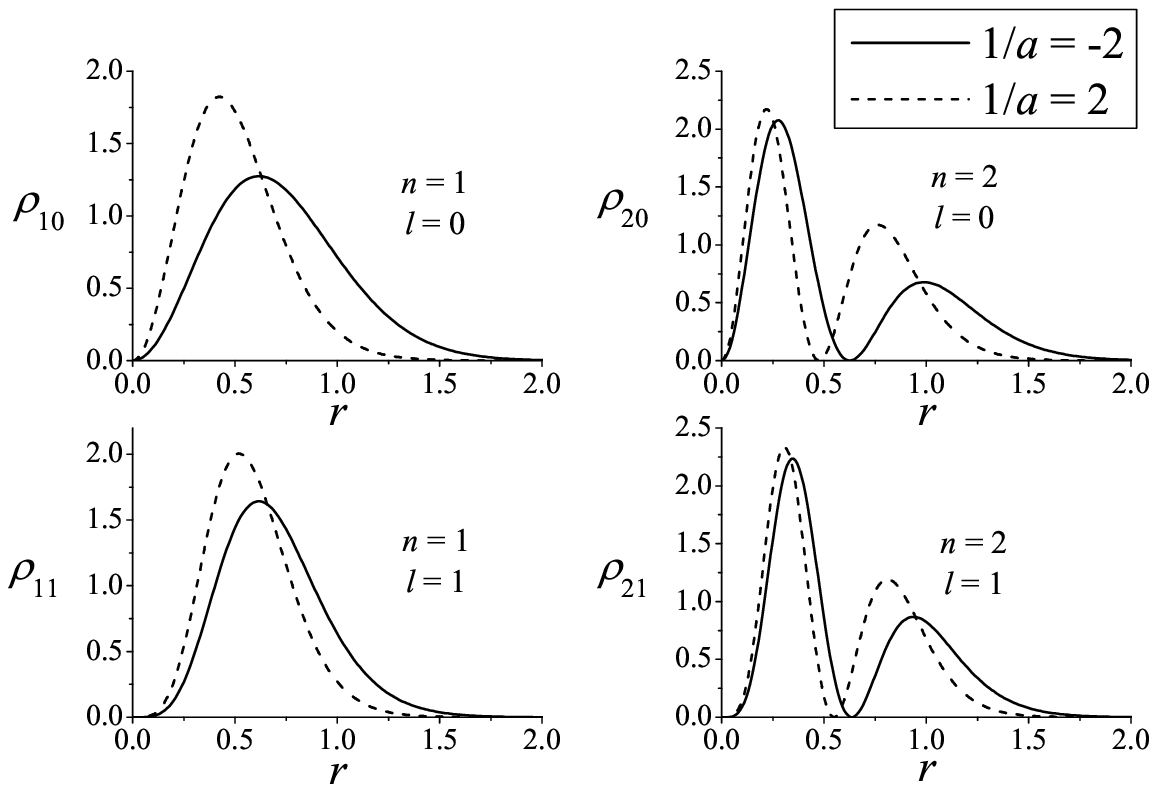}
\caption{Probability density of the radial part of Schmidt modes
$\rho _{nl} = \left| {u_{nl} } \right|^2$, where the $u_{nl}$ is
defined in Eq. (12). The solid line corresponds to the ground
state with $1/a=-2$, while the dash line corresponds to $1/a=2$.
The radial distance $r$ is in units of $\left( {\hbar /\mu \omega
} \right)^{1/2}$.}
\end{figure}

\begin{references}
\bibitem{number} For example, the lowest order van der Waals
interaction has a characteristic range of the order of 50{\rm
\AA}, and the width of typical trap of frequency 100 Hz is of
order 1 $\mu$m.
\end{references}
\end{document}